\documentclass{moriond}
\usepackage{amsmath,amsfonts,amssymb}
\usepackage{graphicx}
\usepackage{mathrsfs}
\usepackage{subcaption}
\usepackage{comment}

\def\be{\begin{equation}}
\def\ee{\end{equation}}
\def\bea{\begin{eqnarray}}
\def\eea{\end{eqnarray}}

\newcommand{\Photo}{\includegraphics[height=35mm]{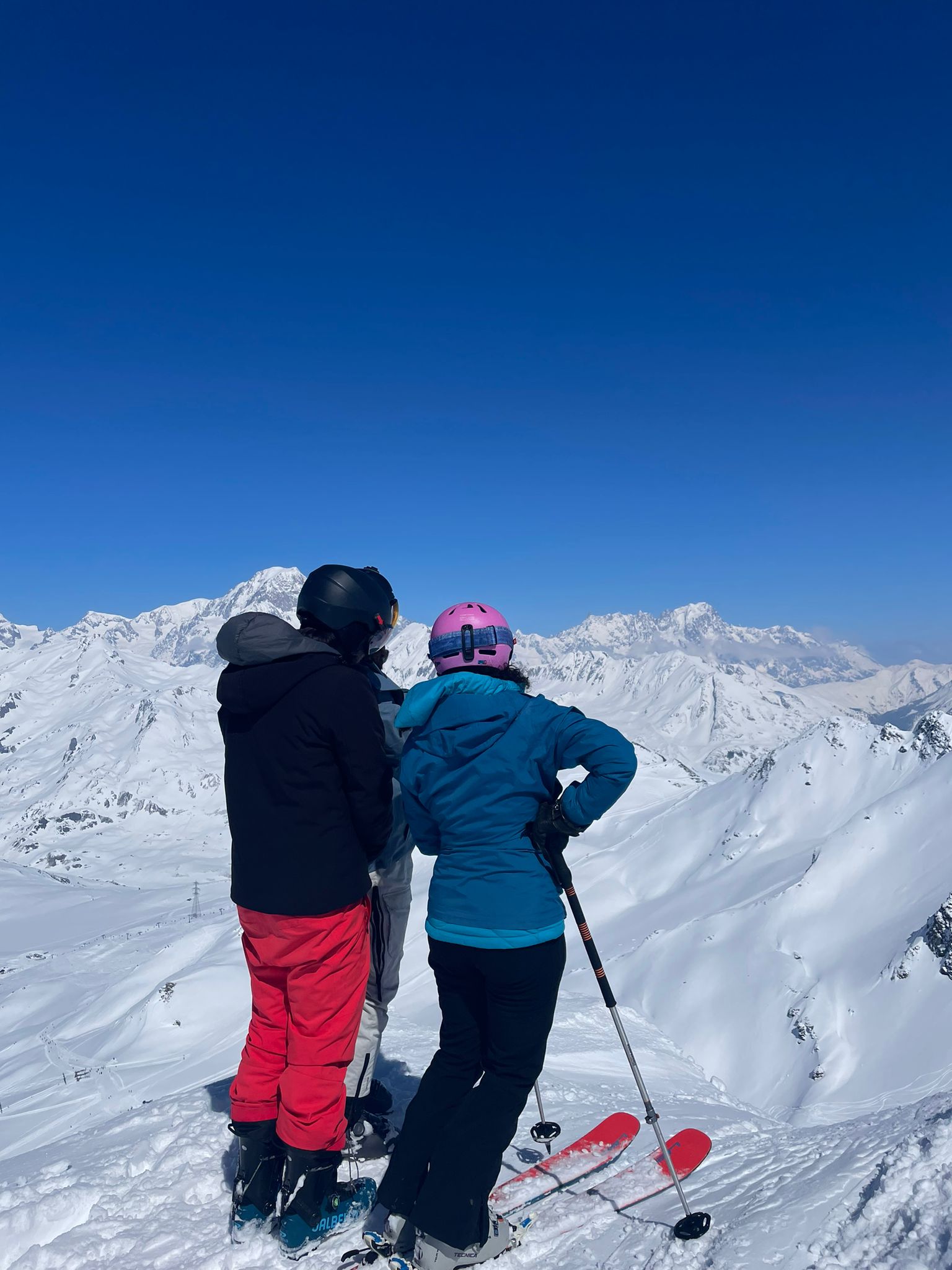}}

\begin{document}
\vspace*{4cm}
\title{Gravitational Wave Memory in Beyond GR Theories}

\author{ Silvia Gasparotto }

\address{CERN, Theoretical Physics Department, Esplanade des Particules 1, Geneva 1211, Switzerland}

\maketitle
\begin{abstract}
Gravitational-wave memory is a low-frequency, non-oscillatory signal that provides a promising probe of strong-field gravity. We present the first computation of memory from full inspiral--merger--ringdown waveforms in a theory beyond GR, focusing on scalar Gauss--Bonnet gravity. We find percent-level deviations from GR, mainly driven by modified merger dynamics, while scalar-induced contributions to tensor memory are strongly suppressed. We found that including memory greatly enhances the mismatch between GR and beyond-GR waveforms, highlighting its potential as a complementary observable for tests of gravity with next-generation detectors.
\end{abstract}

\section{Does space time remember?}

Gravitational-wave (GW) signals from binary black hole (BBH) mergers are typically described as oscillatory chirps that decay once the remnant black hole settles down. However, GW bursts also induce a permanent shift in the spacetime metric, known as \textit{gravitational-wave memory}, first identified in~\cite{Zeldovich:1974gvh,Braginsky:1985vlg,Braginsky:1987kwo}.
Memory has two contributions: \textit{linear memory}, sourced by asymmetric mass ejection, and \textit{nonlinear memory}, generated by the GW energy flux itself~\cite{Favata:2010zu}. We focus on the latter, which is the dominant effect in BBH mergers and is a generic prediction of General Relativity (GR). Also called \textit{displacement memory}, it produces a lasting relative displacement between freely falling observers. This zero-frequency component is linked to asymptotic spacetime symmetries (BMS) and to Weinberg’s soft graviton theorem~\cite{Strominger:2017zoo}.

For an isolated source, the memory is given by the integral of the GW energy flux over the entire emission history:
\begin{equation}
    \label{eq:GWmem1}
     \delta h_{ij}^{\mathrm{TT}}(u) = \frac{4}{d} \int_{-\infty}^{u} \mathrm{d}u' \int \mathrm{d}\Omega \, \frac{\mathrm{d}^2 E^{\mathrm{GW}}}{\mathrm{d}u'\mathrm{d}\Omega} \left[ \frac{n_i n_j}{1 - n_k N^k} \right]^{\mathrm{TT}},
\end{equation}
where $d$ is the source distance, $\vec{n}$ the emission direction, and $\vec{N}$ the line of sight.

The memory signal appears as a step-like feature, with a sharp growth near merger where most of the energy is emitted as shown in Fig.~\ref{fig:memplot}. In the frequency domain, it is dominated by low frequencies, making its detection challenging for current interferometers~\cite{Zosso:2026czc}. Third-generation detectors such as the {Einstein Telescope} and {Cosmic Explorer}, as well as the space-based mission {LISA}, are expected to enable its observation~\cite{Grant:2022bla,Cogez:2026frh,Inchauspe:2024ibs,Gasparotto:2023fcg}.
Therefore, a central question is what information can be extracted from GW memory. Since standard tests of GR rely on parametrized waveform deviations, understanding how memory is modified in theories beyond GR is a necessary step toward using it as an additional probe of gravity.
\section{Memory in theories with additional scalar field: the scalar Gauss-Bonnet case}

\begin{figure}[t]
    \centering
    \begin{subfigure}[b]{0.42\textwidth}
        \centering
        \includegraphics[width=\textwidth]{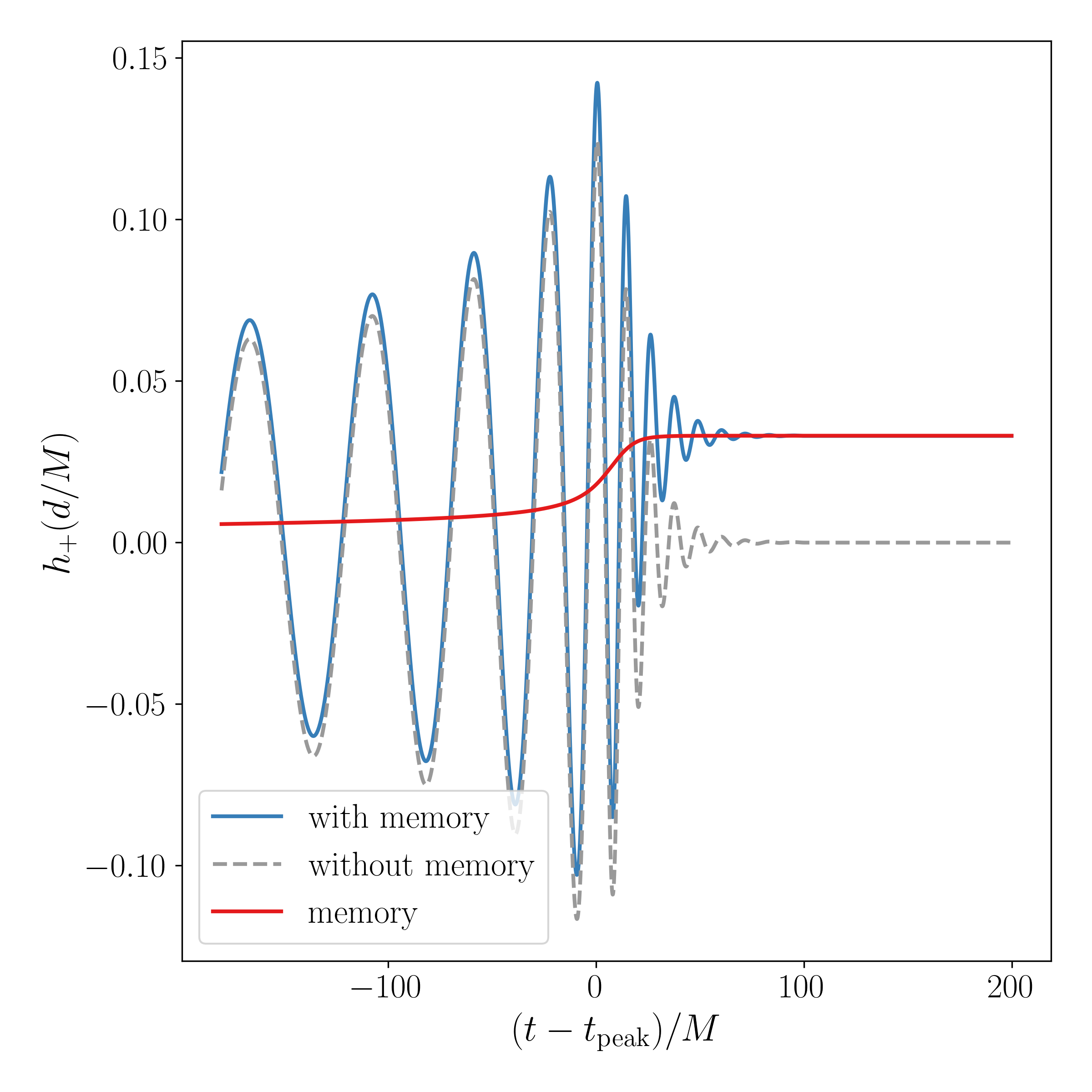}
        
    \end{subfigure}
    \hfill
    \begin{subfigure}[b]{0.52\textwidth}
        \centering
        \includegraphics[width=\textwidth]{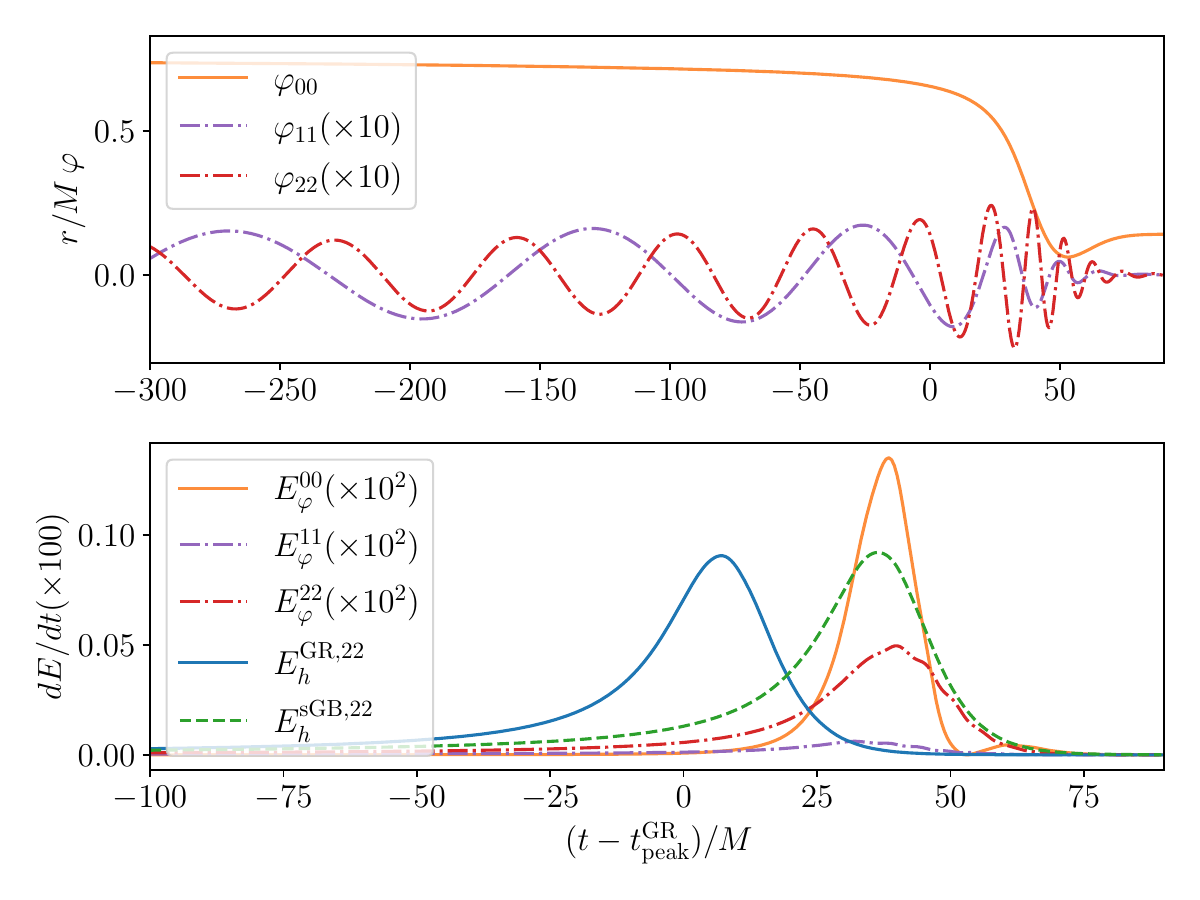}
        
    \end{subfigure}
    \caption{\textit{Left:} GW strain with and without memory for an equal-mass, edge-on GR binary. 
\textit{Right:} Scalar-field waveform and energy fluxes for the same system as in Fig.~\ref{fig:waveforms_sGB}.}
    \label{fig:memplot}
\end{figure}
The theoretical framework for GW memory in modified gravity was developed in Refs.~\cite{Heisenberg:2023prj,Heisenberg:2024cjk}. These works showed that, in metric theories satisfying the Einstein equivalence principle and containing additional massless degrees of freedom, the memory contribution retains the same structural form as in GR, with all radiative degrees of freedom entering through the total energy flux; the general expression is given in Eq.~(90) of Ref.~\cite{Heisenberg:2023prj}. However, while GW memory has been studied in beyond-GR theories at the post-Newtonian level, calculations in the strong-field merger regime, where the observable signal is expected to be largest, have so far been missing. In our work~\cite{Gasparotto:2026bru}, we address this gap by providing the first computation of GW memory from full inspiral--merger--ringdown waveforms in scalar Gauss--Bonnet (sGB) gravity. This theory is a natural setting for such a study, as it contains an additional scalar degree of freedom, constitutes one of the most developed frameworks admitting black-hole solutions with scalar hair, and enables fully nonlinear numerical-relativity simulations in the weak-coupling regime.
The action reads
\begin{equation}\label{eq:ActionsGB}
    S=\,\frac{1}{16\pi}\int d^4x\sqrt{-g}\left(R-\frac{1}{2}(\nabla\Phi)^2+\lambda\, f(\Phi){\mathcal R}^2_{\rm GB}\right) +S_\text{m}[g,\Psi_\text{m}]\,,
\end{equation}
where ${\mathcal R}^2_{\rm GB}$ is the Gauss--Bonnet invariant, $\lambda$ is a coupling constant, and $f(\Phi)$ specifies the theory. We consider both linear coupling, $f(\Phi)=\Phi$, where all black holes carry scalar hair, and nonlinear coupling, e.g.\ $f(\Phi)=\frac{1}{2\beta}(1-e^{-\beta\Phi^2})$, which allows for dynamical scalarization. In the latter case, scalar hair is triggered when the binary reaches sufficiently high curvature, corresponding to a short critical distance.

\begin{figure}
    \centering
    \includegraphics[width=0.95\linewidth]{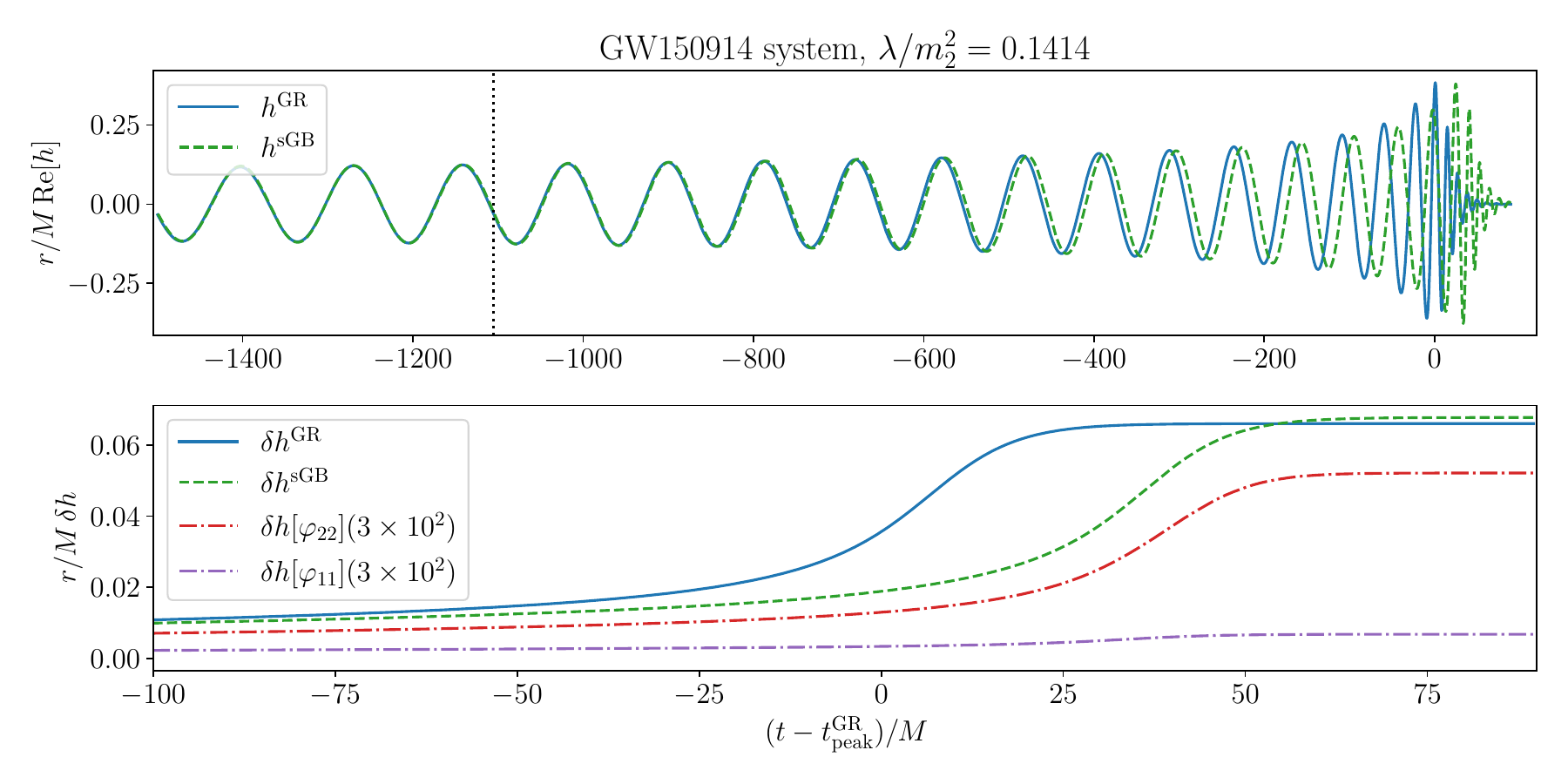}
    \caption{Gravitational waveforms for a near-equal-mass binary in GR and sGB gravity with linear coupling $\lambda/m_2^2=0.1414$ and mass ratio $q=1.221$ (GW150914\-like). The final memory amplitude is $0.0634$ in sGB and $0.0619$ in GR at $r=150M$. The vertical dotted line indicates the end of the alignment interval (see Ref.~\protect\cite{Corman:2025wun}). }
    \label{fig:waveforms_sGB}
\end{figure}

In this theory, deviations from GR can arise from two effects: modifications of the dominant GW signal and an additional contribution to the memory sourced by the scalar-field energy flux, which we refer to as scalar-induced memory.
An example waveform is shown in Fig.~\ref{fig:waveforms_sGB}, based on the simulations of Ref.~\cite{Corman:2025wun} for a GW150914-like system. Comparing GR ($\lambda=0$) and sGB ($\lambda/m_2^2=0.1414$), we find two main effects: a shift in the merger time and an increase in the final memory amplitude, at the level of $\sim 2.5\%$, due to enhanced nonlinear dynamics in sGB. 

The total memory is dominated by the tensor contribution sourced by the $(2,2)$ mode. Scalar radiation, shown in Fig.~\ref{fig:memplot}, contributes subdominantly: most of it is emitted in the monopole, which does not generate tensor memory, while higher multipoles are suppressed. As a result, scalar-induced memory is more than two orders of magnitude smaller than the tensor contribution. 
Further details, including higher mass ratios and dynamical scalarization effects, can be found in Ref~\cite{Gasparotto:2026bru}.

\section{Relevance of the effect for tests of GR}

The deviations in the memory signal discussed above remain small. It is therefore essential to assess their detectability and possible degeneracies with standard waveform parameters. A common measure is the mismatch $\mathcal{M}$, defined from the noise-weighted inner product
\begin{equation}
\langle h_1 | h_2 \rangle = 4\,\mathrm{Re} \int_{f_{\min}}^{f_{\max}} \frac{\tilde{h}_1(f)\,\tilde{h}_2^*(f)}{S_n(f)} \, df \,, 
\quad 
\mathcal{M} = 1- \max_{\Delta t,\,\Delta \phi} \frac{\langle h_1 | h_2 \rangle}{\sqrt{\langle h_1 | h_1 \rangle \langle h_2 | h_2 \rangle}} \, ,
\end{equation}
where the maximization is performed over time and phase shifts.

Given the limited availability of beyond-GR waveforms, we consider variations in the total mass, which primarily rescale the signal. Even in this simplified setting, including memory increases the mismatch between GR and sGB waveforms by more than an order of magnitude. Therefore, the inclusion of the memory enhances the distinguishability between the two waveforms, thereby increasing the observable volume over which such differences can be detected. For a $20\,M_\odot$ binary observed with the Einstein Telescope, this corresponds to sensitivity up to $z\lesssim 0.2$, i.e.\ $\mathcal{O}(0.1\text{--}1)\,\mathrm{yr}^{-1}$ events.

For the system considered, the best-fitting GR waveform has a total mass $\simeq 0.99\,M$. Since lower-mass systems produce smaller memory, this induces a residual $\sim 4\%$ difference in the final memory amplitude, helping to break parameter degeneracies. Such deviations may be observable with next-generation detectors: Ref.~\cite{Goncharov:2023woe} shows that the Einstein Telescope could constrain the memory amplitude at the $\sim 2\%$ level by stacking BBH events.

Although limited by the scarcity of numerical waveforms beyond GR, these results provide a first quantitative target for memory-based tests of gravity. The detection of GW memory would already be a major milestone, and its inclusion in standard analyses can provide complementary information to break degeneracies and enhance tests of GR.

\section*{Acknowledgments}
I would like to acknowledge my collaborators Jann Zosso, Llibert Aresté Saló, Daniela D. Doneva, Stoytcho S. Yazadjiev  for their valuable work and help during the project. 

\section*{References}
\bibliography{moriond}


\end{document}